\documentclass[a4paper,11pt]{article}
\usepackage{jcappub} % for details on the use of the package, please
\usepackage[T1]{fontenc} % if needed

\usepackage{ulem}
\input{colordvi.tex}
\usepackage{aas_macros}

\newcommand{\eqsref}[1]{eq.(\ref{#1})}
\newcommand{\figref}[1]{figure \ref{#1}}
\newcommand{\secref}[1]{section \ref{#1}}

\title{\boldmath Primordial magnetic fields from the string network}

\author[a,1]{Kouichirou Horiguchi,\note{Corresponding author.}}
\author[a,b]{Kiyotomo Ichiki,}
\author[a,b,c]{and Naoshi Sugiyama}

\affiliation[a]{Department of Physics and Astrophysics, Nagoya University,\\ Aichi 464-8602, Japan}
\affiliation[b]{Kobayashi-Masukawa Institute for the Origin of Particles and the Universe, \\Nagoya University, \\Nagoya 464-8602, Japan}
\affiliation[c]{Kavli Institute for the Physics and Mathematics of the Universe (Kavli IPMU), The University of Tokyo, \\Chiba 277-8582, Japan}
\emailAdd{horiguchi.kouichirou@h.mbox.nagoya-u.ac.jp}
\emailAdd{ichiki@a.phys.nagoya-u.ac.jp}
\emailAdd{naoshi@nagoya-u.jp}

\abstract{ Cosmic strings are a type of cosmic defect formed by a
symmetry-breaking phase transition in the early universe.  Individual
strings would have gathered to build a network, and their dynamical
motion would induce scalar--, vector-- and tensor--type perturbations.
In this paper, we focus on the vector mode perturbations arising from
the string network based on the one scale model and calculate the time
evolution and the power spectrum of the associated magnetic fields. We
show that the relative velocity between photon and baryon fluids induced
by the string network can generate magnetic fields over a wide range of
scales based on standard cosmology.  We obtain the magnetic field
spectrum before recombination as
$a^2B(k,z)\sim4\times10^{-16}G\mu/((1+z)/1000)^{4.25}(k/{\rm
Mpc}^{-1})^{3.5}$ Gauss on super-horizon scales, and
$a^2B(k,z)\sim2.4\times10^{-17}G\mu/((1+z)/1000)^{3.5}(k/{\rm
Mpc}^{-1})^{2.5}$ Gauss on sub-horizon scales in co-moving
coordinates. This magnetic field grows up to the end of recombination,
and has a final amplitude of approximately $B\sim10^{-17\sim -18} G\mu$
Gauss at the $k\sim1\ {\rm Mpc}^{-1}$ scale today.  This field might
serve as a seed for cosmological magnetic fields.}

\begin{document}
\maketitle
\flushbottom

%%%%%%%%%%%%%%%%%%%%%%%%%%%%%%%%%%%%%%%%%%%%%%%%%
%%introduction
%%%%%%%%%%%%%%%%%%%%%%%%%%%%%%%%%%%%%%%%%%%%%%%%%
\section{Introduction}
Various observations, such as the cosmic microwave background (CMB) and
the dimming of distant supernovae, support the standard big-bang model,
which has gained a credible position in modern cosmology in recent
years. In standard big-bang cosmology, because of its cooling from a
very hot initial state following adiabatic expansion, the universe has
experienced a number of phase transitions. These phase transitions are
expected to generate topological defects in the universe. The nature of
these resulting defects depends on the type of symmetry breaking caused
by the phase transitions. For example, the topological defects generated
in ${\cal O}(N)$ symmetry-breaking phase transitions are called ${\cal
O}(N)$ global defects and correspond to domain walls, cosmic strings,
monopoles, and textures for ${\cal O}(N=1)$, ${\cal O}(N=2)$, ${\cal
O}(N=3)$, ${\cal O}(N\geq4)$ symmetry breaking, respectively.

Among others, cosmic strings are expected to affect physics at various
scales \cite{2015arXiv150604039V}.  There are a number of phenomena
caused by cosmic strings, for instance, as gravitational signatures,
primordial gravitational waves (PGWs) from cusps and kinks on infinite
strings and string loops
\cite{2000PhRvL..85.3761D,2012PhRvD..86b3503K,2014PhRvD..89b3512B,2010PhRvD..81j3523K},
gravitational lensing by strings (strong lensing
\cite{2007PhRvD..76l3515M} and microlensing \cite{2008MNRAS.384..161K}),
CMB angular power spectra
\cite{1999PhRvD..60h3504P,2012PhRvD..86l3513A}, and so on.  As
non-gravitational signatures, the following are predicted: ultra-high
energy cosmic rays from strings \cite{2000PhR...327..109B} and cusps on
string loops via a scalar field \cite{2010PhRvD..81d3531V}, and radio
bursts from kinks and cusps on strings via the gravitational
Aharonov-Bohm effect \cite{2010PhRvD..81d3503J,2011PhRvD..83d3528S}.

Because of the impact on physics, a number of studies exist that aim to
place constraints on string tension $\mu$ which shows the string energy
scale. As for CMB observations, cosmic strings induce CMB anisotropies
of the order of $\Delta T/T\sim 4\pi G\mu$, where $T$ is the CMB
temperature, $\Delta T$ is its fluctuation, and $G$ is the gravitational
constant. As a result, the recent CMB temperature measurement by the
Planck collaboration provides limits to the energy scale of cosmic
strings \cite{2014A&A...571A..25P}. Details of the limits depend on the
models of the cosmic strings, for instance, $G\mu\leq 3.2\times 10^{-7}$
for Abelian Higgs strings and $G\mu\leq1.5\times10^{-7}$ for Nambu-Goto
strings \cite{2014A&A...571A..25P}.

In this paper, we investigate primordial magnetic fields induced by the
cosmic string network. Previous works about the generation of primordial
magnetic fields from cosmic strings include generation from the motion
of wiggly strings \cite{PhysRevLett.67.1057}, shock waves induced by
cosmic strings \cite{PhysRevD.48.3585,PhysRevD.45.3487}, the dynamical
friction of strings' motion \cite{PhysRevD.51.5946}, and via the
Harrison mechanism in the early universe
\cite{2008PhRvD..77f3517H}. Because of the conservation of vorticity, it
is argued that primordial magnetic fields can hardly be produced by
cosmic defects \cite{2008PhRvD..77f3517H}. To reassess the generation of
primordial magnetic fields from cosmic strings, we focus on the tight
coupling relationship between photon and baryon fluids in the early
universe and the anisotropic stress of photons.  We see that generation
of magnetic fields from the cosmic string network is possible if we
consider up to second order in the tight coupling expansion including
the anisotropic stress of photons.  In order to calculate the time
evolution of cosmic string networks and their associated magnetic
fields, we modified CMBACT \cite{act}; this is the code used to
calculate the evolution of the string network and CMB anisotropies from
the cosmic string network following the "one scale model"
\cite{1985NuPhB.252..227K,PhysRevD.34.3592,Martins:1996jp}.
 
In the next section, we describe the model of the individual strings and
the evolution of the string network. In \secref{sec:mag_gen}, we
investigate magnetic fields generation from the string network by
considering the tight coupling approximation between the photon and
baryon fluids. In \secref{sec:method}, the method which accounts for the
randomness of the strings' initial configuration is given. We then
discuss evolution of the magnetic field spectrum before and after the
recombination epoch in \secref{sec:disc}. Finally, we summarize the
features of the magnetic fields from the cosmic string network in
\secref{sec:sum}. Throughout this paper, we assumed an homogeneous and
isotropic expanding universe consistent with the $\Lambda$-CDM model as
the background metric.  We fixed the cosmological parameters to
$h=0.73$, $\Omega_mh^2=0.127$, $\Omega_bh^2=0.0223$, and $N_\nu=3.04$,
where $H_0=100h{\rm km}/{\rm s}/{\rm Mpc}$ is the Hubble constant,
$\Omega_m$ and $\Omega_b$ are the density parameters of matter and
baryon, and $N_\nu$ is the number of massless neutrinos.

%%%%%%%%%%%%%%%%%%%%%%%%%%%%%%%%%%%%%%%%%%%%%%%%%
%%model of strings
%%%%%%%%%%%%%%%%%%%%%%%%%%%%%%%%%%%%%%%%%%%%%%%%%
\section{Model of the strings and their evolution}
\label{string_model_evolve} In this section, we review the evolution of
individual strings and the string network. We first introduce the
energy--momentum tensor of the individual strings. Moreover, by considering
the evolution of the separations, motions, and decays of the strings, we
calculate the energy--momentum tensor of the string network. Here we
assume a homogeneous and isotropic expanding universe described by the
FLRW metric given by
\begin{equation}
\label{eq:flrw}
ds^2 = a(\tau^2)(d\eta^2-d\vec{x}^2).
\end{equation}

\subsection{Wiggly string}
Here we introduce the action and the energy stress tensor of a wiggly
cosmic string on the string worldsheet following
\cite{1999PhRvD..60h3504P}.  We define the string worldsheet
$(\zeta_1,\zeta_2)$ in four dimensional space-time, where $\zeta_1=\eta$
is the conformal time defined in \eqsref{eq:flrw}. Using these
conditions, we can calculate the action density as
\begin{eqnarray}
S&=&-\mu\int_{\Sigma} dA\\
&=&-\mu\int_{\Sigma} d^2\zeta \left[-{\rm det}_{ab}\left(\gamma^{ab}\right)\right]^{\frac{1}{2}},
\end{eqnarray}
where
\begin{equation}
\gamma^{ab}=g_{\mu\nu}\frac{\partial x^\mu}{\partial \zeta_a}\frac{\partial x^\nu}{\partial \zeta_b}
\end{equation}
is the metric on the string worldsheet. Here $\mu$ is the string energy
density per unit length.  From the definition, we can write the
energy--stress tensor of the string as
\begin{eqnarray}
\label{eq:emt}
T^{\mu\nu}=\frac{\mu}{\sqrt{-g}}\int_{\Sigma} d^2\zeta\ \left\{\epsilon\dot{x}^\mu \dot{x}^\nu-\epsilon^{-1}x'^\mu x'^\nu \right\}\delta^{(4)}(y-x(\zeta)),
\end{eqnarray}
with
\begin{equation}
\epsilon=\sqrt{\frac{\vec{x}'^2}{1-\dot{\vec{x}}^2}},
\end{equation}
where dots and primes represent derivatives with respect to the
conformal time and $\zeta_2$, respectively, and we have chosen the gauge
as $\dot{x}^\mu x'_\mu=0$. Following \cite{PhysRevD.41.3869}, the string
tension $T$ and the linear energy density $U$ can be defined by
\begin{eqnarray}
\label{eq:emt_wtu}
\sqrt{-g}T^{\mu\nu}(y)=\int_{\Sigma} d^2\zeta\ \sqrt{-\gamma}\left\{Uu^\mu u^\nu-Tv^\mu v^\nu \right\}\delta^{(4)}(y-x(\zeta)),
\end{eqnarray}
where 
\begin{equation}
\label{eq:uvprop1}
u^\mu u_\mu=-v^\mu v_\mu=1,\quad u^\mu v_\mu =0,
\end{equation}
\begin{equation}
\label{eq:uvprop2}
(u^\mu v^\rho -v^\mu u^\rho)(u_\rho v_\nu -v_\rho u_\nu)=u^\mu u_\nu-v^\mu v_\nu= \eta^\mu_\nu,
\end{equation}
\begin{equation}
\label{eq:uvprop3}
\eta^{\mu\nu}=\gamma^{ab}x^\mu_{,a}x^\nu_{,b},
\end{equation}
and
\begin{equation}
\label{eq:uv_to_x}
u^\mu=\frac{\sqrt{\epsilon}\dot{x}^\mu}{(-\gamma)^{1/4}},\quad v^\mu=\frac{x'^\mu}{\sqrt{\epsilon}(-\gamma)^{1/4}}.
\end{equation}
We can easily check that $u^\mu$ and $v^\mu$ satisfy \eqsref{eq:uvprop1}
$\sim$ \eqsref{eq:uvprop3}.  Substituting \eqsref{eq:uv_to_x} into
\eqsref{eq:emt_wtu} and comparing with \eqsref{eq:emt}, we can see that
\begin{equation}
U=T=\mu.
\end{equation}
This is the equation of state for a Nambu--Goto string.  In practice, in
lattice simulations of cosmic strings, the strings are macroscopically
straight, but they have wiggles from a microscopic viewpoint
\cite{PhysRevD.41.2408,gibbons1990formation}. Even though the cosmic
strings are macroscopic and straight on the cosmological scale
($\sim{\rm Mpc}$), their macroscopic equations of state are affected by the
microscopic wigglinesses. The equation of state for the wiggly string,
smoothing out the small scale structures, is shown in
\cite{PhysRevD.41.3869,PhysRevD.41.3038}. In this case, we employ effective
values of the string tension $\tilde{T}$ and the linear energy density
$\tilde{U}$, with which the equations of states for the wiggly string are denoted
as
\begin{equation}
\label{eq:eos_eff}
\tilde{T} = \frac{\mu}{\alpha},\quad \tilde{U} = \alpha\mu.
\end{equation}
Here $\alpha$ is the ``wiggliness parameter'', which is a function of
time and the string coordinate $\zeta$. The evolution of $\alpha$ is
estimated in
\cite{1999PhRvD..60h3504P,PhysRevD.41.2408,PhysRevLett.64.119}, which
show that $\alpha \sim 1.9$ in the radiation dominated era and
$\alpha\sim1.5$ in the matter dominated era. In the late time universe,
when the cosmological constant dominates, wiggliness is smoothed out by
the exponential expansion of the universe and $\alpha$ becomes equal to
unity. Substituting the effective values of \eqsref{eq:eos_eff} into
\eqsref{eq:emt_wtu}, we can write the effective energy--momentum tensor
of a wiggly string $\tilde{T}^{\mu\nu}$ as
\begin{eqnarray}
\tilde{T}^{\mu\nu}(y)&=&\frac{1}{\sqrt{-g}}\int_{\Sigma} d^2\zeta\ \sqrt{-\gamma}\left\{\tilde{U}u^\mu u^\nu-\tilde{T}v^\mu v^\nu \right\}\delta^{(4)}(y-x(\zeta))\\
&=&\frac{\mu}{\sqrt{-g}}\int_{\Sigma} d^2\zeta\ \left\{\epsilon\alpha\dot{x}^\mu \dot{x}^\nu-\frac{x'^\mu x'^\nu}{\epsilon\alpha}\right\}\delta^{(4)}(y-x(\zeta)).
\end{eqnarray}
If we consider magnetic field generation by the string network, the
wiggliness affects the magnetic field spectrum only through a constant
factor $\alpha^2$. Here we set $\alpha = 1$ for simplicity, which
represents straight strings.

\subsection{Evolution of the  string network}
Following the one-scale model
\cite{1985NuPhB.252..227K,PhysRevD.34.3592,Martins:1996jp}, the network
of cosmic strings can be characterized by a single parameter, the
correlation length $L$, which is defined as
\begin{equation}
L^2 = \frac{\mu}{\rho_{\rm string}},
\end{equation}
where $\rho_{\rm string}$ is the energy density of the cosmic string. To
simplify the expressions, we introduce the comoving correlation length $l = L/a$.  From
the energy conservation law and the equation of motion of the string, we
can obtain the evolution equations of the string network as
\cite{1985NuPhB.252..227K,PhysRevD.34.3592,Martins:1996jp}:
\begin{eqnarray}
\frac{dl}{d\eta}&=&{\cal H}lv^2+\frac12 \tilde{c}v,\\
\frac{dv}{d\eta}&=&(1-v^2)\left(\frac{\tilde{k}}{l}-2{\cal H}v\right),
\end{eqnarray}
where ${\cal H}=\dot{a}/a$ is the conformal Hubble parameter,
\begin{equation}
v=\sqrt{\frac{\int d\sigma \epsilon\vec{\dot{x}}^2}{\int d\sigma \epsilon}},\nonumber
\end{equation}
is the string's rms velocity, $\tilde{c}$ is the loop--chopping efficiency, 
\begin{equation}
\tilde{k}=\frac{\int d\sigma \epsilon (1-\vec{\dot{x}}^2)\vec{\dot{x}}\cdot\hat{u}}{v(1-v^2)\int d\sigma \epsilon},\nonumber
\end{equation}
is the effective curvature of the strings, and $\hat{u}$ is the unit
vector of the curvature radius vector of the string.
Here we consider the reduction of the energy density by the expansion of
the universe and the decrease of the total length of infinite strings in
the Hubble horizon.  The total length of infinite strings decreases
because infinite strings are chopped due to their collisions, making
string loops that immediately decay by radiating gravitational
waves from their peakedness \cite{PhysRevD.64.064008}. The loop--chopping
efficiency $\tilde{c}$ represents the rate at which strings become
loops.  In general, $\tilde{c}$ is a function of time as shown in
\cite{1999PhRvD..60h3504P,Martins:1996jp}; however its value does not
vary much. 
%constants.
  In the radiation dominated era, $\tilde{c}=c_r = 0.23$, and
in the matter dominated era, $\tilde{c} = c_m=0.18$.

We describe the Fourier transform of the energy--momentum tensor of an
individual string as
\begin{eqnarray}
\Theta^{\mu\nu}(\vec{k},\eta)&=&\int d^3x e^{i\vec{k}\cdot\vec{x}}T^{\mu\nu}(x)\nonumber\\
&=&\int_{-l/2}^{l/2} d\zeta e^{i\vec{k}\cdot \vec{X}}\left[\epsilon\alpha \dot{X}^\mu\dot{X}^\nu -\frac{X'^\mu X'^\nu}{\epsilon \alpha}\right],
\end{eqnarray}
where the four vector $X^\mu(\zeta,\eta)=(\eta,\vec{X})$ denotes the
coordinates in which the string segment exists. We can represent the vector
as
\begin{equation}
X^0=\eta,\quad \vec{X}=\vec{x}_0+\zeta\hat{X}'+v\eta \hat{\dot{X}},
\end{equation}
where $\vec{x}_0$ is a random vector that denotes the initial
position of the string's mass center, and $\hat{X}'$ and $\hat{\dot{X}}$ are
random unit vectors that fulfill $\hat{X}'\cdot\hat{\dot{X}}=0$. In
the expression for the energy--momentum tensor of the string, the initial
position coordinate $\vec{x}_0$ appears only as a phase in the inner
product with $\vec{k}$. We can therefore deal with
$\vec{k}\cdot\vec{x}_0$ as a random initial phase $\phi_0:[0,2\pi]$.
Because individual strings have their own initial positions, aligned
directions, and velocity directions, we write those of the $m$-th string
as $\vec{x}_0^m,\hat{X}'^m$ and $\hat{\dot{X}}^m$. Summing up our
description of strings, 
we obtain the total energy momentum tensor of the string network.
Fourier transform of the total energy momentum tensor of the string network $\Theta_{\mu\nu}(\vec{k},\tau)$ is given by \cite{1999PhRvD..60h3504P},
\begin{equation}
\Theta_{\mu\nu}(\vec{k},\eta) = \sum_{m=1}^{N_0}\Theta^m_{\mu\nu}(\vec{k},\eta) T^{\rm off}(\eta,\eta_m),
\end{equation}
where $N_0$ is the initial number of the strings, $\Theta^m_{\mu\nu}(\vec{k},\tau)$ is the energy momentum tensor of the individual strings, $\eta_m$ is the time of decay for the $m$-th string, and $T^{\rm off}$ is the smoothing function of the decay. We adopt the functional form of $T^{\rm off}$ from \cite{1999PhRvD..60h3504P,PhysRevD.59.023508} that is given by
\begin{eqnarray}
\label{eq:decay_func}
T^{\rm off}(\eta,\eta_m)&=&\left\{ \begin{array}{ll}
1,\quad&(\eta<f\eta_m)\\
1/2+(x^3-3x)/4,\quad&(f\eta_m<\eta<\eta_m)\\
0,\quad&(\eta_m<\eta),\\
\end{array}\right.\\
x&=&2\frac{{\rm ln}(f\eta_m/\eta)}{{\rm ln}f}-1.
\end{eqnarray}
Here $0< f < 1$ is a parameter which controls the speed of the strings decay. We
 fix this value to $f=0.5$. Considering \eqsref{eq:decay_func}
for individual infinite strings, we take account of the decrease in
the number of infinite strings by their decay into loops due to their
intersections. Because of their random initial positions and directions
of motion, the decay time for each string $\eta_m$ is fixed randomly.

%%%%%%%%%%%%%%%%%%%%%%%%%%%%%%%%%%%%%%%%%%%%%%%%%
%%magnetic fields
%%%%%%%%%%%%%%%%%%%%%%%%%%%%%%%%%%%%%%%%%%%%%%%%%

\section{Magnetic fields}
\label{sec:mag_gen}
Infinite strings can be the source of magnetic fields on large scales
around the recombination era. In this section, we estimate the amplitude
of the magnetic fields produced by the network made of infinite strings.
\subsection{Vector mode perturbation}
Here we take the Poisson gauge,
\begin{equation}
ds^2=a^2(\eta)(-(1+2\psi)d\eta^2+2w_id\eta dx^i+[(1-2\phi)\delta_{ij}+h_{ij}]dx^idx^j).
\end{equation}
In the same way as \cite{2015JCAP...04..007H}, using the vector
projector tensor
\begin{equation}
{\cal O}_{ij}^{(\lambda)}(\hat{k})=\frac{i\lambda}{\sqrt{2}}(\hat{k}_ie^{(\lambda)}_j(\hat{k})+\hat{k}_je^{(\lambda)}_i(\hat{k})),
\end{equation}
we can denote the vector mode part of $h_{ij}$ directly as 
\begin{equation}
h_{ij}=\sum_{\lambda=\pm1}h_V^{(\lambda)}{\cal O}_{ij}^{(\lambda)}.
\end{equation}
The evolution equation of the vorticity $\sigma=\dot{h}_V/k$ is given
based on the Einstein equation by
\begin{equation}
\label{eq:vort}
\dot{\sigma}^{(\lambda)}+2{\cal H}\sigma^{(\lambda)} = 8 \pi Ga^2\Pi^{(\lambda)}/k.
\end{equation}
Here $\lambda = \pm$ is the index of polarization, and
$\Pi=\Theta_{ij}^{\rm tot}(k,\tau){\cal O}_{ij}$ is the total
anisotropic stress in the vector mode. In this paper, we assume the
infinite strings to be the sources of the vector mode perturbations.

In the vector mode, the Euler equation for the baryon fluid is given by
 \begin{equation}
 \label{eq:BV}%\eqsref{eq:BV}
 \dot{v}_b-\dot{\sigma}+{\cal H}(v_b-\sigma)=R\dot{\tau}(v_{\gamma}-v_b),
 \end{equation}
where $v_\gamma$ and $v_b$ are the velocities of photon and baryon
fluids, respectively, $\rho_\gamma$ and $\rho_b$ are the energy
densities of photon and baryon fluids, respectively,
$R=4\rho_\gamma/3\rho_b$ is the photon-baryon ratio, $\dot{\tau}
=a\sigma_Tn_e$ is the opacity of the Thomson scattering, $\sigma_T$ is
the Thomson scattering cross section, and $n_e$ is the electron number
density. The vector mode Boltzmann equations for the photon fluid are
given by
\begin{eqnarray}
\label{eq:PV}%\eqsref{eq:PV}
\dot{v}_\gamma-\dot{\sigma}+\frac{k}{8}\Pi_\gamma&=&-\dot{\tau}(v_{\gamma}-v_b),\\
\dot{\Pi}_{\gamma}+\frac{8}{5}kI_3-\frac{8}{5}kv_{\gamma}&=&-\dot{\tau}\left(\frac{9}{10}\Pi_{\gamma}-\frac{9}{5}E_2\right),\nonumber\\
\label{eq:Pig}%\eqsref{eq:Pig}
\\
\label{eq:IH}%\eqsref{eq:IH}
\dot{I}_l+k\frac{l}{2l+1}\left(\frac{l+2}{l+1}I_{l+1}-I_{l-1}\right)&=&-\dot{\tau}I_l \quad\quad(l\geq3),
\end{eqnarray}
\begin{eqnarray}
& &\dot{E}_l+\frac{(l+3)(l+2)l(l-1)}{(l+1)^3(2l+1)}kE_{l+1}-\frac{l}{2l+1}kE_{l-1}\nonumber\\
& &\quad\quad\quad\quad\quad\quad\quad\quad\quad\quad\quad\quad=-\dot{\tau}\left(E_l-\frac{2}{15}\xi\delta_{l2}\right)+\frac{2}{l(l+1)}kB_l,\quad\nonumber\\
\label{eq:EH}%\eqsref{eq:EH}
\\
& &\dot{B}_l+\frac{(l+3)(l+2)l(l-1)}{(l+1)^3(2l+1)}kB_{l+1}-\frac{l}{2l+1}kB_{l-1}=-\frac{2}{l(l+1)}kE_l, \quad\nonumber\\
\label{eq:BH}%\eqsref{eq:BH}
\end{eqnarray}
where $\Pi_\gamma=3I_2$ is the anisotropic stress of the photon fluid,
$I_l$ is the $l$-th order moment of the intensity, $E_l$ and $B_l$ are
the $l$-th order moments of the polarization, and $\xi=3I_2/4+9E_2/2$
\cite{2004PhRvD..70d3518L}.

As shown in \cite{2015JCAP...04..007H}, the topological defects induce
$v_\gamma$ and $v_b$ from the vorticity $\sigma$. Then, the relative
velocity between the photon and baryon fluids, $v_\gamma-v_b$, plays the
main role in exciting the magnetic fields
\cite{2005PhRvL..95l1301T,2006astro.ph..1243T}.  The evolution of the
relative velocity and thus the evolution of the associated magnetic
fields are driven by the strength of the coupling between the photon and
baryon fluids. Therefore, the magnetic fields evolve differently
before and after the epoch of recombination. Before recombination, a 
tight-coupling approximation can be applied to describe their evolution.
However, after recombination, we need to solve the baryon fluid equation
\eqsref{eq:BV} and the full Boltzmann equations
\eqsref{eq:PV}$\sim$\eqsref{eq:BH}. We do this numerically.

\subsection{Tight-coupling approximation}
In the early universe before recombination, photon and baryon fluids are
 tightly coupled to each other because of the frequent Thomson
 scattering. In that epoch, the opacity of the Thomson scattering
 $\dot{\tau}$ was very large and the tight-coupling parameter
 $k/\dot{\tau}$ takes a very small value
 $(k/\dot{\tau}\ll1)$. Therefore, we can expand the Boltzmann and the
 Einstein equations with the tight coupling parameter. This expansion is
 called the tight-coupling approximation (TCA). In
 \cite{PhysRevD.85.043009}, the authors considered magnetic field
 generation with no external source, and  used the first order
 approximation for the photon's anisotropic stress $\Pi_\gamma^{(1)}$
 and the second order approximation for the relative velocity between
 photon and baryon fluids $v^{(2)}_\gamma-v^{(2)}_b$. In the case where
 there is an external source, higher order TCA, such as
 $\Pi_\gamma^{(2)}$ and $v_\gamma^{(3)}-v_b^{(3)}$ are needed
 \cite{2015JCAP...04..007H}. In this paper, we consider the string
 network as the external source and we need to consider the higher order
 approximation.

Here we consider the TCA up to the third order in the conformal
Newtonian gauge.  In the TCA, we expand the relative velocity as $\delta
v=v_\gamma-v_b=0+\delta v^{(1)}+\delta v^{(2)}+\delta v^{(3)}+...$ ,
where $\delta v^{(n)} \propto (k/\dot{\tau})^n$ is the $n$-th order
expansion. Following \cite{2015JCAP...04..007H}, we find the TCA up to
second order for $\Pi_\gamma$ and up to third order for $\delta v$ as,
\begin{eqnarray}
\label{eq:nlsm_pig3}%\eqsref{eq:nlsm_pig3}
\Pi_{\gamma}^{(1)}&=&\frac{32}{15}\left(\frac{k}{\dot{\tau}}\right)v_{\gamma}^{(0)},
\nonumber \\
\Pi_{\gamma}^{(2)}&=&\frac{32}{15}\left(\frac{k}{\dot{\tau}}\right)v_{\gamma}^{(1)}+\frac{176}{45}\left(\frac{k}{\dot{\tau}}\right)^2\frac{1}{k}\left[\frac{\ddot{\tau}}{\dot{\tau}}v_{\gamma}^{(0)}-\dot{v}_{\gamma}^{(0)}\right],
\end{eqnarray}
\begin{eqnarray}
\delta v^{(1)}&=&\left(\frac{k}{\dot{\tau}}\right)\frac{\cal H}{(1+R)k}(v_{\gamma}^{(0)}-\sigma^{(0)}),\\
\label{eq:nlsm_dv3}%\eqsref{eq:nlsm_dv3}
\delta v^{(2)}&=&\left(\frac{k}{\dot{\tau}}\right)\frac{\cal H}{(1+R)k}(v_{\gamma}^{(1)}-\sigma^{(1)})-\frac{4}{15}\left(\frac{k}{\dot{\tau}}\right)^2\frac{1}{1+R}v_{\gamma}^{(0)}\nonumber\\
& &-\left(\frac{k}{\dot{\tau}}\right)^2\frac{{\cal H}(v_{\gamma}^{(0)}-\sigma^{(0)})}{(1+R)^2k^2}\left(\frac{{\cal H}R}{1+R}+\frac{\dot{\cal H}}{\cal H}+{\cal H}+\frac{\dot{v}_{\gamma}^{(0)}-\dot{\sigma}^{(0)}}{v_{\gamma}^{(0)}-\sigma^{(0)}}-\frac{\ddot{\tau}}{\dot{\tau}}\right),\\
\delta v^{(3)}& = &\left(\frac{k}{\dot{\tau}}\right)\frac{\cal H}{(1+R)k}(v_{\gamma}^{(2)}-\sigma^{(2)})-\frac{4}{15}\left(\frac{k}{\dot{\tau}}\right)^2\frac{1}{1+R}v_{\gamma}^{(1)}\nonumber\\
& &-\left(\frac{k}{\dot{\tau}}\right)^2\frac{{\cal H}(v_{\gamma}^{(1)}-\sigma^{(1)})}{(1+R)^2k^2}\left(\frac{{\cal H}R}{1+R}+\frac{\dot{\cal H}}{\cal H}+{\cal H}+\frac{\dot{v}_{\gamma}^{(1)}-\dot{\sigma}^{(1)}}{v_{\gamma}^{(1)}-\sigma^{(1)}}-\frac{\ddot{\tau}}{\dot{\tau}}\right)\nonumber\\
& &+\frac{4}{15}\left(\frac{k}{\dot{\tau}}\right)^3\frac{\cal H}{(1+R)^2k}v_\gamma^{(0)}\nonumber\\
&&-\frac{2}{45k}\left(\frac{k}{\dot{\tau}}\right)^3\frac{1}{(1+R)^2}\left[(23+11R)\frac{\ddot{\tau}}{\dot{\tau}}v_\gamma^{(0)}-(17+11R)\dot{v}_\gamma^{(0)}-\frac{6v_\gamma^{(0)}{\cal H}R}{1+R}\right]. 
\end{eqnarray}
In this paper, we assume that the strings are the only source of
vorticity. In this case, the fact that $v^{(0)}=\sigma^{(0)}$ plays the
most important role. Because of this, the first order TCA of the
relative velocity is given as $\delta v^{(1)} = 0$. Therefore, $\delta
v^{(2)}\sim(k/\dot{\tau})^2\sigma/(1+R)$ is the leading order of the
TCA.  In our numerical calculation, evolution equations are switched
from the TCA to the full Boltzmann equations at the epoch of
recombination. At this time, we need to calculate an accurate relative
velocity $\delta v^{(2)}=v_\gamma^{(2)}-v_b^{(2)}$ via the Boltzmann
equations. The junction conditions for $v_\gamma^{(2)}$ and $v_b^{(2)}$
at recombination are given by the following equations,
\begin{eqnarray}
\dot{v}_b^{(2)}-\dot{\sigma}^{(2)}+{\cal H}(v_b^{(2)}-\sigma^{(2)})&=&R\dot{\tau}\delta v^{(3)}_{\rm TCA},\\
\dot{v}_\gamma^{(2)}-\dot{\sigma}^{(2)}+\frac{k}{8}\Pi_\gamma^{(2)}&=&-\dot{\tau}\delta v^{(3)}_{\rm TCA}.
\end{eqnarray}
Therefore we need $\delta v^{(3)}$ for an accurate calculation of
$\delta v^{(2)}$ at the switching time \cite{2015JCAP...04..007H} from
the TCA to the full Boltzmann equations.

\subsection{Magnetic field generation}
The relative velocity between the photon and baryon fluids can induce
magnetic fields \cite{2005PhRvL..95l1301T,2006astro.ph..1243T}. In the
early universe, electrons move together with photons because of the
frequent Thomson scattering.  Because this scattering
separates electrons from photons, electric fields are induced and
their rotations generate magnetic fields via the Maxwell equations. The
evolution equation of the magnetic fields is given by
\cite{2005PhRvL..95l1301T,2006astro.ph..1243T,2007astro.ph..1329I},
\begin{equation}
\label{eq:mag_evolve}
\frac{1}{a}\frac{\rm d}{{\rm d} \eta} (a^2B^i)=\frac{4\sigma_T\rho_\gamma a}{3e}\epsilon^{ijk}\partial_k(v_{\gamma j}-v_{bj})~,
\end{equation}
where $e$ is the elementary charge and $\epsilon^{ijk}$ is the
Levi-Civita tensor. We can obtain the magnetic field spectrum by
integrating \eqsref{eq:mag_evolve} in Fourier space as
\begin{eqnarray}
\label{eq:mag_spec}
\left<a^4B^i(\vec{k},\eta)B_i^*(\vec{k}',\eta)\right>&=&\left(\frac{4\sigma_T}{3e}\right)^2(\delta^{jl}\delta^{km}-\delta^{jm}\delta^{kl})k_kk'_m\left<\int _0^{\eta} {\rm d}\eta'a^2(\eta')\rho_{\gamma}(\eta')\delta v_j(\vec{k},\eta')\right.\nonumber\\
& &\left.\times\int _0^{\eta} {\rm d}\eta''a^2(\eta'')\rho_{\gamma}(\eta'')\delta v_l^*(\vec{k}',\eta'')\right>.
\end{eqnarray}
The ensemble average of the relative velocity can be written as 
\begin{eqnarray}
\label{eq:v_spec}%\eqsref{eq:MF_nlsm_IS}
\left<
\delta v_j(\vec{k},\eta')\delta v^*_l(\vec{k}',\eta'')\right>
&=&P_{jl}(\hat{k}) \delta v (k,\eta') \delta v (k,\eta'')(2\pi)^3\delta(\vec{k}-\vec{k}'),\\
\label{eq:projection}%\eqsref{eq:Projection}
P_{jl}(\hat{k})&=&\delta_{jl}-\hat{k}_j\hat{k}_l.
\end{eqnarray}
Substituting \eqsref{eq:v_spec} and \eqsref{eq:projection} into
\eqsref{eq:mag_spec}, we obtain the correlation function of the magnetic
fields as \cite{PhysRevD.85.043009}
\begin{equation}
\label{eq:spec}
\left<B^i(\vec{k},\eta)B_i(\vec{k}',\eta)\right>=(2\pi)^3S_B(k,\eta)\delta^{(3)}(\vec{k}-\vec{k}'),
\end{equation}
where
\begin{equation}
\label{eq:mag_pow_spec}
a^4(\eta)\frac{k^3}{2\pi^2}S_B(k,\eta)=2\frac{k^3}{2\pi^2}\left(\frac{4\sigma_T}{3e}\right)^2k^2 \left[\int _0^{\eta} {\rm d}\eta'\ a^2(\eta')\rho_{\gamma}(\eta')\delta v(k,\eta')\right]^2.
\end{equation}
The source of this spectrum is $\delta v$, which is driven by $\sigma$.  

%%%%%%%%%%%%%%%%%%%%%%%%%%%%%%%%%%%%%%%%%%%%%%%%%
%%method
%%%%%%%%%%%%%%%%%%%%%%%%%%%%%%%%%%%%%%%%%%%%%%%%%

%%K.I editing here
\section{Method}
\label{sec:method} Once the initial configuration of the string network
and its evolution are fixed, the spectrum of the magnetic fields can be
calculated as shown in \secref{sec:mag_gen}.

However, the string network has a random initial configuration, and
individual strings decay at random.  To see the statistical properties
of the generated magnetic fields, we need to average out their
randomness. 
For that,
 we prepare a number of realizations for the string network and
calculate the power spectrum \eqsref{eq:mag_pow_spec} under each
realization.

In practice, we realize the magnetic fields by repeatedly following the three steps listed below using CMBACT\cite{act}:\\\\
$1.$ Set a random initial configuration of the string network.\\
$2.$ Compute the evolution of the energy--momentum tensor of the string network by considering random decay of the strings.\\
$3.$ Calculate the magnetic field spectrum.\\\\
In the $m$-th realization, the $m$-th power spectrum $S_B^m(k,\eta)$ in
\eqsref{eq:spec} is calculated. Moreover, we can obtain the averaged power spectrum $S_B^{\rm ave}(k,\eta)$ as
\begin{equation}
\label{eq:ave_spec}
S_B^{\rm ave}(k,\eta)=\frac{1}{N_{\rm r}}\sum^{N_{\rm r}}_{m=1}S_B^m(k,\eta),
\end{equation}
where $N_{\rm r}$ is the number of realizations, and we fix $N_{\rm r}=100$ in this paper.
Because each $S_B^m$ has an initial configuration and evolution, the
averaged power spectrum, $S_B^{\rm ave}(k,\eta)$, can not be divided
into an initial power spectrum and common transfer functions. Therefore,
to see the statistical properties of the generated magnetic fields, we
need to calculate a number of spectra under different realizations and average them
out as \eqsref{eq:ave_spec}. 

\begin{figure}[!h]
\centering
\includegraphics[width=.75\textwidth]{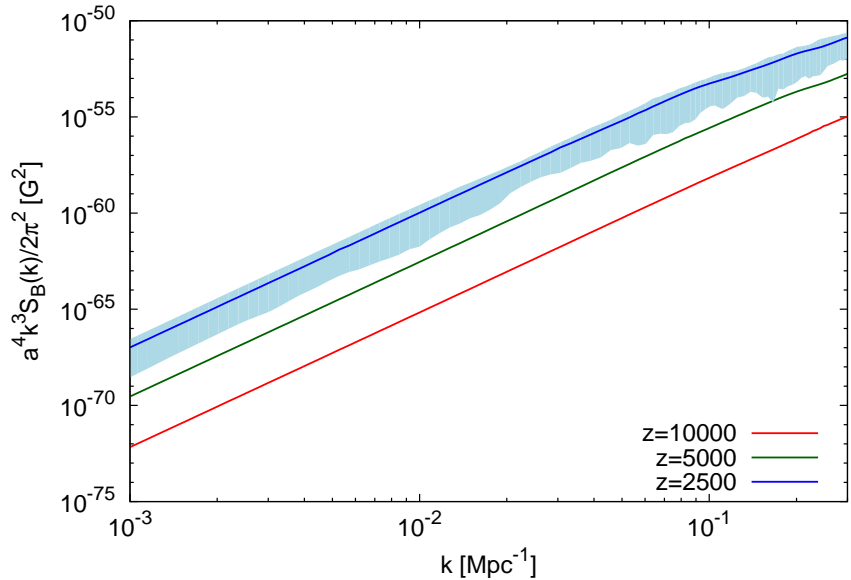}
\caption{\label{fig:mag_spec_tca_v} 
Power spectra of magnetic fields at $z=10,000$ (red solid line), $5,000$
 (green solid line), and $2,500$ (blue solid line) from the infinite string network with $G\mu=1.1\times10^{-6}$. We averaged $100$ realizations and the light-blue region is the $68\%$ confidence interval for the $z=2,500$ case. Under these redshifts, the TCA is valid.
 }
\end{figure}

\begin{figure}[!h]
\centering
\includegraphics[width=.75\textwidth]{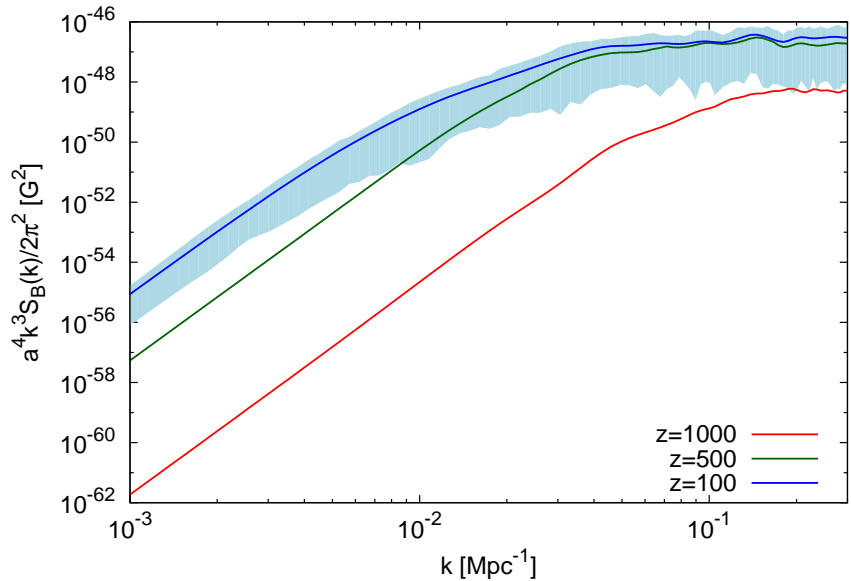}
\caption{\label{fig:mag_spec_tca_inv} 
Same as \figref{fig:mag_spec_tca_v}, but at $z=1,000,\ 500,\ 100$. In these epochs, the TCA is invalid.
 }
\end{figure}

According to the above argument and calculating \eqsref{eq:mag_pow_spec}
numerically, we can obtain the power spectrum of the magnetic fields from
the string network as shown in \figref{fig:mag_spec_tca_v} and
\figref{fig:mag_spec_tca_inv}. Figure \ref{fig:mag_spec_tca_v} shows
magnetic field spectra before recombination when the TCA can be applied,
whereas \figref{fig:mag_spec_tca_inv} shows the spectra after
recombination when the TCA is invalid.

%we  can't divide S_B into initial spectrum and transfer function
%Some attentions and sequence for calculating power spectrum of the cosmic string network using Cmbact   . we need to ....

%%%%%%%%%%%%%%%%%%%%%%%%%%%%%%%%%%%%%%%%%%%%%%%%%
%%discussion for  magnetic fields spectrum
%%%%%%%%%%%%%%%%%%%%%%%%%%%%%%%%%%%%%%%%%%%%%%%%%

\section{Result \& Discussion}
\label{sec:disc}

In this section, we will give an analytical interpretation of the
magnetic field spectrum arising from the string network. Here to understand the
behavior of the spectrum, we investigate the evolution of magnetic
fields separately before and after recombination.

\subsection{Before recombination}
 Before recombination, Thomson scattering between photons and electrons
 occurs frequently; therefore, we can use the TCA. Focusing on the
 vector mode that is responsible for the generation of  magnetic fields,
 all of the perturbations are induced by vorticity $\sigma$ from the
 string network. 
This assumption leads the first order TCA of the relative velocity to be
 zero $\delta v^{(1)}=0$; therefore, the second order TCA of the relative
 velocity $\delta v^{(2)}$ is the leading order of $\delta v$
 \cite{2015JCAP...04..007H}, 
\begin{eqnarray}
\label{eq:src_tcav}
\delta v\approx \delta v^{(2)}\propto\left(\frac{k}{\dot{\tau}}\right)^2\frac{v_\gamma^{(0)}}{1+R}=\left(\frac{k}{\dot{\tau}}\right)^2\frac{\sigma^{(0)}}{1+R}
=\left\{ \begin{array}{ll}
k^2a^5\sigma^{(0)}\quad&({\rm for\ rad.\ dominated\ era})\\
k^2a^4\sigma^{(0)}\quad&({\rm for\ mat.\ dominated\ era}).
\end{array}\right.
\end{eqnarray}
 Sourced by this relative velocity, magnetic fields are generated via
 \eqsref{eq:mag_pow_spec} in individual realizations. Here we define the
 source function of the magnetic fields $F_s(k,\eta)$ as 
 \begin{equation}
 \label{eq:src_func}
F_s(k,\eta)\equiv\int _0^{\eta} {\rm d}\eta'\ a^2(\eta')\rho_{\gamma}(\eta')\delta v(k,\eta').
\end{equation}
%This takes on time evolution of magnetic fields in the comoving coordinate. 
Assuming $\sigma\propto a^\nu$, and integrating \eqsref{eq:vort}, we can
find $\nu\geq-2$ once the anisotropic stress $\Pi$ from the string
network arises. Then, substituting $\sigma\propto a^\nu$ and
\eqsref{eq:src_tcav} into \eqsref{eq:src_func}, we find that the source
function $F_s(k,\eta)$ always increases with time. This
means that the co-moving magnetic fields induced in this era are always
increasing. 

Following the above argument, we obtain a spectrum of the magnetic fields
$S_B^m(k,\eta)$, which grows in time for each realization.  
Averaging  these spectra in the manner explained in \secref{sec:method},
we calculate the averaged magnetic field power spectrum as shown in
\figref{fig:mag_spec_tca_v}. From  \figref{fig:mag_spec_tca_v}, we can
find approximate expressions for the magnetic field spectrum. On
super-horizon scales, the expression is given by 
 \begin{equation}
 a^4(\eta)\frac{k^3}{2\pi^2}S_B(k,z)\approx1.6\times10^{-31}(G\mu)^2\left(\frac{1+z}{1000}\right)^{-8.5}\left(\frac{k}{{\rm Mpc}^{-1}}\right)^7\quad[{\rm G}^2],%1.57==>1.6
 \end{equation}
and on sub-horizon scales,
\begin{equation}
\label{eq:mag_spec_sub}
 a^4(\eta)\frac{k^3}{2\pi^2}S_B(k,z)\approx5.8\times10^{-34}(G\mu)^2\left(\frac{1+z}{1000}\right)^{-7}\left(\frac{k}{{\rm Mpc}^{-1}}\right)^5\quad[{\rm G}^2].%5.7851==>5.8
 \end{equation}
On super horizon scales, we find that the wavenumber dependence is the
same as that of the magnetic field spectrum from texture
\cite{2015JCAP...04..007H} and second order density perturbations
\cite{2007astro.ph..1329I,2015PhRvD..91l3510S}, but slightly different
from the magnetic fields generated in Einstein-aether gravity
$\left<B^2_{\rm EA}(k)\right>\propto  k^{8}$
\cite{2013PhRvD..87j4025S}. 
The power spectrum of anisotropic stress arising from the string network is
shown in \figref{fig:pi_flat}. On sub-horizon scales, 
we find the power spectrum of magnetic fields as
$k^3S_B(k) \propto k^5$. To understand this scale dependence, we show the power
spectrum of anisotropic stress arising from the string network in \figref{fig:pi_flat}.3. On
sub-horizon scales,
the spectrum shows 
$\Pi\propto k^{-1}$. From the equations \eqsref{eq:vort} and \eqsref{eq:nlsm_dv3} we
find the relations $\sigma \propto \Pi/k$ and $\delta v \propto k^2
\sigma$, which imply that $k^3 S_B(k)\propto k^5$ from
\eqsref{eq:mag_pow_spec}. The wavenumber dependence differs from 
the other models which are given by $\left<B^2_{\rm NLSM}(k)\right>\propto  k^{1}$ \cite{2015JCAP...04..007H}, $\left<B^2_{\rm rec}(k)\right>\propto  k^{2}$ or $k^{2/3}$ \cite{2007astro.ph..1329I} and $\left<B^2_{\rm EA}(k)\right>\propto  k^{2}$ or $k^{-2}$ \cite{2013PhRvD..87j4025S}.

On smaller scales before recombination,  $\delta v$ begins to decay at
the Silk damping scale. Here,  the co-moving spectrum
can be written as  \eqsref{eq:mag_spec_sub} $\left<a^4B^2(k,a)\right>
\propto k^5a^{-7}$ and the Silk scale can be written as $k_{\rm
Silk}\propto a^{-3/2}$ \cite{1995ApJ...444..489H}. In the same way as
\cite{2012PhRvD..85d3009I}, we obtain the power spectrum of magnetic
fields at scales  smaller than the Silk scale as $\left<a^4B^2_{\rm
Silk}(k)\right>\propto k^{1/3}$. This spectrum continues to the horizon
scale of electron positron pair annihilation, $k\sim10^{5}{\rm
Mpc}^{-1}$, where the mechanism of magnetic field generation
considered in this paper ceases to function \cite{2012PhRvD..85d3009I}. 

\subsection{After recombination}
Because the TCA was valid before recombination, we only needed the
expression of $\delta v$ up to $\delta v^{(2)}$ for the source of
the magnetic fields. 
However, around recombination, as the number density of free electrons
decreases, the frequency of  the Thomson scattering between photons and
electrons decreases. Moreover the photons and electrons gradually
decouple. Then, the baryon fluid
becomes less able to catch up with photon fluid and the TCA breaks down. After this switching epoch, 
to obtain $\delta v $, we need to calculate the full Boltzmann-Einstein system \eqsref{eq:vort}$\sim$\eqsref{eq:BH}. 
In general, it is difficult to solve the Boltzmann equations and see the evolution of $\delta v$ analytically. However, on super-horizon scales,  we can estimate $\delta v$ using the condition $k\eta\ll1$. By integrating \eqsref{eq:BV} and \eqsref{eq:PV}, baryon and photon fluids velocities can be denoted as
 \begin{eqnarray}
\label{eq:vb_int}
v_b(k,\eta)&=&\sigma(k,\eta)+\frac{1}{a(\eta)}\int^\eta d\eta'a(\eta')\dot{\tau}(\eta')R(\eta')\delta v(k,\eta')\\
&=&\sigma(k,\eta)+R(\eta)\int^\eta d\eta'\dot{\tau}(\eta')\delta v(k,\eta'),\\
 \label{eq:vg_int}
 v_\gamma(k,\eta)&=&\sigma(k,\eta)-\frac{k}{8}\int^\eta d\eta'\Pi_\gamma(k,\eta')-\int^\eta d\eta'\dot{\tau}(\eta')\delta v(k,\eta').
 \end{eqnarray}
From \eqsref{eq:vb_int}$-$\eqsref{eq:vg_int}, we obtain the differential equation for $\delta v$,
\begin{equation}
\dot{\delta v}\simeq -\frac{k}{8}\Pi_\gamma(k,\eta)-\dot{\tau}(\eta)\delta v(k,\eta'),
\end{equation}
where we neglected the second term on the RHS of \eqsref{eq:vb_int} since $R(\eta)=4\rho_\gamma/3\rho_b\ll1$ after recombination. We solve it to  obtain the form of $\delta v(k,\eta)$
\begin{equation}
\delta v(k,\eta) \simeq -\frac{k}{8}e^{-\tau(\eta)}\int^\eta d\eta'\Pi_\gamma(k,\eta')e^{\tau(\eta')}.
\end{equation}
Writing the anisotropic stress of photons from \eqsref{eq:Pig} as,
\begin{equation}
\label{eq:pig_int}
\Pi_\gamma(k\eta)\sim \frac{8}{5}k\int^\eta d\eta'v_\gamma(k,\eta'),
\end{equation}
we can estimate $\delta v$ using $v_\gamma$ as
\begin{equation}
\label{eq:dv_int}
\delta v(k,\eta) \sim -{k^2}e^{-\tau(\eta)}\int^\eta d\eta'e^{\tau(\eta')}\int^{\eta'}d\eta''v_\gamma(k,\eta'').
\end{equation}
Substituting \eqsref{eq:pig_int} and \eqsref{eq:dv_int} into \eqsref{eq:vg_int}, we can see that $v_\gamma(k,\eta)=\sigma(k,\eta)+{\cal O}((k\eta)^2)$. Because of this, $\delta v$ on super-horizon scales should be
\begin{equation}
\label{eq:dv_int_sig}
\delta v(k,\eta) \sim -k^2e^{-\tau(\eta)}\int^\eta d\eta'e^{\tau(\eta')}\int^{\eta'}d\eta''\sigma(k,\eta'').
\end{equation}
After vorticity $\sigma$ becomes source free, $\sigma$ evolves as
$\sigma \propto a^{-2}$ and the source function \eqsref{eq:src_func}
becomes constant. Then, the evolution of  the magnetic fields  finishes at
super-horizon scales. 

On sub-horizon scales, the same argument as that for the super-horizon
holds true, and we can see the same relationship between $\delta v$ and
$v_\gamma$ as in \eqsref{eq:dv_int}. The main difference in this case is
the effects of the higher order terms in $k\eta$. On sub-horizon
scales $(k\eta\geq1)$, the photon fluid velocity $v_\gamma$ evolves
following not the first but the second and third terms on the RHS of
\eqsref{eq:vg_int} (higher order terms in $k\eta$).
%The second term affects as a friction term and the third term affects as a source term. 
Subsequently, after the recombination epoch, the third term vanishes and
the evolution of $v_\gamma$ follows the free-streaming solution. 
Then, the conformal time dependence of \eqsref{eq:dv_int} is up to
$\delta v \propto \eta^2$, and the generation of magnetic fields finishes. 

\begin{figure}[!h]
\centering
\includegraphics[width=.75\textwidth]{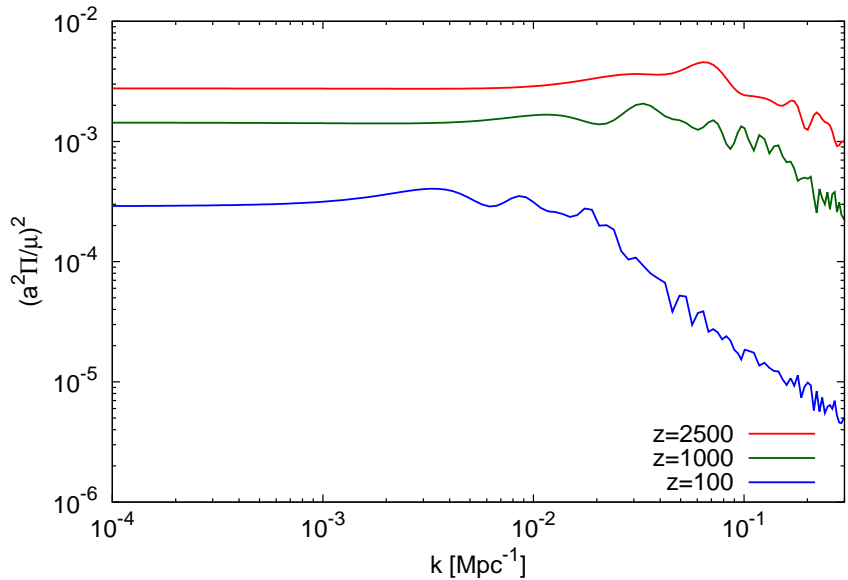}
\caption{\label{fig:pi_flat} 
Wavenumber dependence of the anisotropic stress at $z=2500$ (red solid line), $1000$ (green solid line), $100$ (blue solid line) from the infinite string network.
 }
\end{figure}

In each realization, magnetic fields are induced by this mechanism. As
before, because the evolution of magnetic fields varies in realizations,
we need to take the realization average as in \secref{sec:method}.  The averaged
magnetic field power spectrum after recombination is given in
\figref{fig:mag_spec_tca_inv}. We can see that the anisotropic stress
induced by the string network is independent of the wavenumber $k$ on
the 
super-horizon scale from \figref{fig:pi_flat}. Using \eqsref{eq:vort},
\eqsref{eq:mag_pow_spec} and \eqsref{eq:dv_int_sig}, the wavenumber
power on the super-horizon scale is  the same as that before recombination,
$\left<B^2(k)\right>\propto k^7$. The expression of the magnetic field
spectrum today can be written as 
\begin{equation}
 a^4(\eta)\frac{k^3}{2\pi^2}S_B(k,z)\approx7\times10^{-23}(G\mu)^2\left(\frac{k}{{\rm Mpc}^{-1}}\right)^7\quad[{\rm G}^2],
 \end{equation}
on super-horizon scales and,
\begin{equation}
\label{eq:mag_spec_sub}
 a^4(\eta)\frac{k^3}{2\pi^2}S_B(k,z)\approx2.5\times10^{-35}(G\mu)^2\quad[{\rm G}^2],
 \end{equation}
on sub-horizon scales 
at $z=100$ as shown in \figref{fig:mag_spec_tca_inv}.

 %plus1
 Finally let us discuss implications of these magnetic fields. If such magnetic fields existed in the early universe, strong magnetic
 amplification would occur in the accretion shocks of primordial gases
 during structure formation of the universe.  Then they could provide
 extra pressure and suppress the fragmentation of gas clumps, supporting
 the formation of massive protostars and super massive black holes
 \cite{2014MNRAS.440.1551L}. 
 %plus2
 Moreover, they would affect the hyperfine structure of neutral hydrogens
 in primordial gases and might be observed via the anisotropic power
 spectrum of the brightness temperature of the $21$-cm line with future
 surveys as discussed in \cite{2014arXiv1410.2250V}.

\section{Conclusion}
\label{sec:sum}
 In this paper, we estimated the magnetic field spectrum from the cosmic
 string network. First, we calculated the evolution of the cosmic string
 network under the process of the one scale model and its energy
 momentum tensor using CMBACT \cite{act}. Then, we solved the
 Boltzmann-Einstein system to obtain the relative velocity between the
 photon and baryon fluids using  the tight coupling approximation and
 saw that the leading order of  TCA for $\delta v$ was ${\cal
 O}((k/\dot{\tau})^2)\sigma$ before recombination. Finally, we obtained
 the power spectrum of the magnetic fields via \eqsref{eq:mag_pow_spec},
 before recombination, as
 $a^2\sqrt{B^2(k,z)}\sim4\times10^{-16}G\mu/((1+z)/1000)^{4.25}(k/{\rm
 Mpc}^{-1})^{3.5}$ Gauss on super-horizon scales, and
 $a^2\sqrt{B^2(k,z)}\sim2.4\times10^{-17}G\mu/((1+z)/1000)^{3.5}(k/{\rm
 Mpc}^{-1})^{2.5}$ Gauss on sub-horizon scales in co-moving
 coordinates. On scales smaller than the Silk damping scale, the spectrum
 could be calculated as $a^2\sqrt{B^2(k)}\propto k^{1/6}$. After
 recombination, the spectrum was driven by the evolution of vorticity on
 super-horizon scales and the coupling between photon and baryon fluids
 on sub-horizon scales. When the recombination epoch came to an end,
 the evolution of magnetic fields also ceased. The magnetic field
 spectrum today is $a^2\sqrt{B^2(k,z)}\sim2\times10^{-11}G\mu(k/{\rm
 Mpc}^{-1})^{3.5}$ Gauss on super-horizon scales and
 $a^2\sqrt{B^2(k,z)}\sim5\times10^{-17}G\mu$ Gauss on sub-horizon
 scales. 

\section*{Acknowledgement}
This work is supported in part by a Grant--in--Aid for JSPS Research under Grant No.15J05029 (K.H.) and a Grant--in--Aid for JSPS Grant--in--Aid for Scientific Research under Grant No. 24340048 (K.I.), 25287057 (N.S.) and 15H05890 (N.S.).

\bibliography{ref}
\end{document}